\documentclass[a4paper]{coupled2021}

\usepackage{graphicx}
\usepackage{amsmath}
\usepackage{amsfonts}
\usepackage{amssymb}
\usepackage{hyperref}

\title{Combining Coupled Skorokhod SDEs and Lattice Gas Frameworks for Multi-fidelity Modelling of Complex Behavioral Systems}

\author{Thi Kim Thoa Thieu$^{*}$, Roderick Melnik$^{*}$$^{\dag}$}

\heading{Thoa Thieu and Roderick Melnik}

\address{$^{*}$M3AI Laboratory, MS2Discovery Interdisciplinary Research Institute\\
Wilfrid Laurier University, Waterloo, ON, Canada N2L 3C5
\and
$^{\dag}$BCAM - Basque Center for Applied Mathematics, E-48009, Bilbao, Spain \and
Email: \{tthieu, rmelnik\}@wlu.ca}


\abstract{
To model reliably behavioral systems with complex bio-social interactions, accounting for uncertainty quantification, is critical for many application areas. However, in terms of the mathematical formulation of the corresponding problems, one of the major challenges is coming from the fact that corresponding stochastic processes should, in most cases, be considered in bounded domains, possibly with obstacles. This has been known for a long time and yet, very little has been done for the quantification of uncertainties in modelling complex behavioral systems described by such stochastic processes. In this paper, we address this challenge by considering a coupled system of Skorokhod-type
stochastic differential equations (SDEs) describing interactions between active and passive participants of a mixed-population group. In developing a multi-fidelity modelling methodology for such behavioral systems, we combine low- and high-fidelity results
obtained from (a) the solution of the underlying coupled system of SDEs and (b) simulations with a
statistical-mechanics-based lattice gas model, where we employ a kinetic Monte Carlo procedure. Furthermore, we provide representative numerical examples of healthcare systems, subject to an epidemic, where the main focus in our considerations is given to an interacting particle system of asymptomatic and susceptible populations.
 }

\begin{document}
%\maketitle

\section{Introduction}

The modelling of complex behavioral systems with bio-social interactions offers many challenging questions to science and technology in general. In developing more realistic models, the uncertainty quantification plays an important role for the setting of mathematical formulation of the problems. Uncertainties coming from different sources are ubiquitous in the modelling of behavioral
systems with bio-social interactions \cite{Bellomo2017}. 
%The second challenge is that the overall dynamics of biological factors is restricted to a bounded domain such as barrier options problems, near ground particle diffusion with deposition, queueing theory and neuronal models. For the class of models with bounded domain, the boundary condition must be specified. The boundary conditions can be reflecting, absorbing or even mixed boundary condition.
In terms of the mathematical formulation of the
corresponding problems, one of the major challenges is coming from the fact that
corresponding stochastic processes describing such systems should, in most cases, be considered in bounded domains,
possibly with obstacles \cite{Skorohod1961,Zhang2016,Buisson2020}. Hence, for the class of stochastic models with bounded domains, the appropriate boundary conditions must be specified. The boundary conditions can be, for example, reflecting, absorbing or even mixed boundary conditions. Recent developments reported in \cite{Thieu2020-3, Briand2020} bring us to the study of a system of Skorokhod-type SDEs modelling the interactions between active and passive populations.
%corresponding problems, one of the major challenges is coming from the fact that
%corresponding stochastic processes describing such systems should, in most cases, be considered in bounded domains,
%possibly with obstacles \cite{Skorohod1961,Zhang2016,Buisson2020}. Hence, for the class of stochastic models with bounded domains, the appropriate boundary conditions must be specified. The boundary conditions can be, for example, reflecting, absorbing or even mixed boundary conditions. Recent developments reported in \cite{Thieu2020-3, Briand2020} bring us to the study of a system of Skorokhod-type SDEs modelling the interactions between active and passive populations.
% The reflecting stochastic dynamics have many applications in queueing network models, mathematical finance, neuroscience and operations research in general.
% Therefore, we pay our attention to the study of a system of Skorokhod-type of stochastic differential equations modeling the interactions between active and passive populations. The reflecting stochastic dynamics have many applications in queueing network models, mathematical finance, neuroscience and operations research in general.
%Operations research that includes queueing theory models is one of important applications of such reflecting stochastic dynamics.
Many important operations research problems, including those arising in queueing theory, require the models describing reflecting stochastic dynamics. Moreover,
 this class of queueing theory models is important in many applied fields, including the emergency medical aid systems \cite{Meares2020}, passenger services in air terminals, road traffic and intelligent transportation, as well as various types of other systems and processes, represented as some operations on material, informational or other flows. On the other hand, the topics of SARS-CoV-2, which is an ongoing global pandemic, have recently attracted the interest of the scientific community. One of the main challenges lies with the development of such mathematical models where the dynamics of social factors can be predicted with high reliability. Furthermore, a suitable model can help prevent the large number of asymptomatic cases that effectively spread the
 virus, which can cause a medical emergency to very susceptible individuals. A number of relevant results are available on these SARS-CoV-2 epidemic topics in particular, as well as on complex systems in general. Specifically, the authors in \cite{Schweitzer2019} have provided an agent-based framework to model self-organizing systems at different levels of organization, physical, biological and social, using the same
dynamic approach.  A novel stochastic model aiming to account for the unique transmission dynamics
of COVID-19 and capture the effects of intervention measures has been proposed in \cite{Zhang2020}. An agent-based
model for simulations of infection transmission through bio-social stochastic dynamics has been investigated in \cite{Tadic2020}. The same authors also extended this agent-based modelling framework to shed light on the role of asymptomatic
hosts and unravel the interplay between the biological and social factors of these nonlinear stochastic processes at high temporal resolution in \cite{Tadic2021}. In \cite{Subramanian2021}, the authors have considered a model for quantifying asymptomatic infection and transmission of COVID-19 by using observed cases, serology, and testing capacity. The presence of asymptomatic hosts in the population is a major challenge for controlling the epidemics given the fact that a new host can contract the virus with a delay. 
The asymptomatic virus carriers can infect a susceptible individual, who, depending on the susceptibility, may or may not develop symptoms. Similarly, the viruses produced by pre-symptomatic hosts can lead to asymptomatic as well as symptomatic cases. 
Taking the inspiration from the study of a system of Skorokhod-type SDEs together with applications in epidemic problems, our representative example here is pertinent to the modelling of healthcare systems, subject to an epidemic, where we have to deal with distinct behaviors caused by the interactions between asymptomatic and susceptible populations.
% by using the combination of coupled Skorhokhod SDE and lattice gas frameworks. 

%In the communication among individuals,

Therefore, in what follows, we consider a system of SDEs of Skorohod type aimed at modelling the dynamics of active-passive populations. In particular, we show that a model of queueing theory converges to a system of reflected SDEs via a limit theorem. Furthermore, we develop a multi-fidelity modelling methodology (e.g. in \cite{Peherstorfer2017, Peherstorfer2018-2}) for our behavioral system. The analysis of this model is carried out by combining low- and high-fidelity results obtained from the solution of the underlying coupled system of SDEs and from the simulations with a statistical-mechanics-based lattice gas model, where we employ a kinetic Monte Carlo
procedure. We provide details of the models developed here, as well as several representative numerical examples  of healthcare systems, subject to an epidemic, where we consider an interacting particle system of asymptomatic and susceptible populations. 

\section{Model description}\label{model}

In this section, we consider a general queueing theory model for active and passive particles. The geometry is a square lattice $\Lambda:=\{1,\dots,L\}\times\{1,\dots,L\}\subset \mathbb{Z}^2$ of side 
length $L$, where $L$ is an odd positive integer number. $\Lambda$ will be referred in this context as {\em room}.  
An element $x=(x_1,x_2)$ of the room $\Lambda$ is called \emph{site} 
or \emph{cell}.
Two sites $x,y\in\Lambda$ are said \emph{nearest neighbors} if and only 
if $|x-y|=1$. The \textquotedblleft door\textquotedblright  \ will be understood here as a set made of $\omega$ pairwise adjacent sites, with $\omega$ and odd positive integer smaller than $L$, located on the top row of the room $\Lambda$ symmetric with respect to its median column. This mimics the presence of an exit door on the top row of the room $\Lambda$. The number $\omega$ will be called \textit{width} of the exit. In addition, $N_A$ is the total number of active particles, $N_P$ is the total number of passive particles with $N:=N_A+N_P$ and $N_A, N_P, N \in \mathbb{N}$. 

The motivation for the above consideration becomes clear if we assume that the exit door is a service station located on the top row of the geometry, and there are $\omega$ service gates at the service station. Every active particle's target is to reach these service gates and to leave the geometry immediately after that, while the passive particles cannot leave the room due to invalid access to the service gates. 

We introduce a special combination of SDEs with
jumps (see, e.g., \cite{Situ2005}) for the following active-passive populations.
\subsection{Active population}
For $i \in \{1,\ldots, N_A\}$ and $t \in [0,T]$, let $X_{A_i}$ denote the position of particle $i$ belonging to the active population at time $t$. Suppose that $f_{A_i}: [0, T]\times \mathbb{R}^2 \longrightarrow \mathbb{R}^2$, $g_{A_i}: [0, T]\times \mathbb{R}^2 \longrightarrow \mathbb{R}^2 \times \mathbb{R}^2$. We consider the dynamics of active particles that is governed by  
\begin{align}\label{active}
\begin{cases}
d X_{A_i}(t) = f_{A_i}(X_{A_i}(t),X_{P_k}(t),t)dt + g_{A_i}(X_{A_i}(t),X_{P_k}(t),t)dW_{A_i}(t) +  d\Phi_{A_i}(t),\\
X_{A_{i}}(0) = X_{A_{i0}}.
\end{cases}
\end{align}
%where $X_{A_{i0}}$ represents the initial configuration of active pedestrians inside $L$.
In other words, the active population of the system is described by the following process:
\begin{align}
X_{A_i}^\alpha(t) &= X_{A_i}^\alpha(0) + \int_{0}^{t}f_{A_i}(X_{A_i}^{\alpha}(s),X_{P_k}^\mu(s),s)ds \nonumber\\&+ \int_{0}^{t}g_{A_i}(X_{A_i}^{\alpha}(s),X_{P_k}^\mu(s),s)dW_{A_i}(s) + \Phi_{A_i}^{\alpha}(t) \nonumber\\
&= X_{A_i}^\alpha(0) + Y_{A_i}^{\alpha}(t) + \Phi_{A_i}^{\alpha}(t),
\end{align}
where $$Y_{A_i}^{\alpha}(t) := \int_{0}^{t}f_{A_i}(X_{A_i}^{\alpha}(s),X_{P_k}^\mu(s),s)ds + \int_{0}^{t}g_{A_i}(X_{A_i}^{\alpha}(s),X_{P_k}^\mu(s),s)dW_{A_i}(s).$$
Here, $(W_{A_i})_i$ is a sequence of independent standard Brownian motions, while $Y_{A_i}^{\alpha}$
can be interpreted as the arrival times of active particles, and $\Phi_{A_i}^{\alpha}(t)$ is the cumulative lost service capacity over $[0,T]$.

\subsection{Passive population}

For $k \in \{1,\ldots, N_P\}$ and $t \in [0,T]$, let $X_{P_k}$ denote the position of particle $k$ belonging at time $t$ to the passive population. Suppose that $f_{P_k}: [0, T]\times \mathbb{R}^2 \longrightarrow \mathbb{R}^2$, $g_{P_k}: [0, T]\times \mathbb{R}^2 \longrightarrow \mathbb{R}^2 \times \mathbb{R}^2$. The dynamics of the passive particles is described  here by a system of stochastic differential equations as follows:
\begin{align}\label{passive}
\begin{cases}
d X_{P_k}(t) = f_{P_k}(X_{A_i}(t),X_{P_k}(t),t)dt + g_{P_k}(X_{A_i}(t),X_{P_k}(t),t)dW_{P_k}(t) +  d\Phi_{P_k}(t),\\
X_{P_{k}}(0) = X_{P_{i0}}.
\end{cases}
\end{align}
%where $X_{P_{k0}}$ represents the initial configuration of passive pedestrians inside $L$.
In other words, the passive population of the system is described by the following process:

\begin{align}
X_{P_k}^\mu(t) &= X_{P_k}^\mu(0) + \int_{0}^{t}f_{P_k}(X_{A_i}^{\alpha}(s),X_{P_k}^\mu(s),s)ds \nonumber\\&+ \int_{0}^{t}g_{P_k}(X_{A_i}^{\alpha}(s),X_{P_k}^\mu(s),s)dW_{P_k}(s) + \Phi_{P_k}^{\mu}(t) \nonumber\\
&= X_{P_j}^\mu(0) + Y_{P_j}^{\mu}(t) + \Phi_{P_j}^{\mu}(t),
\end{align}
where $$Y_{P_j}^{\mu}(t) := \int_{0}^{t}f_{P_k}(X_{A_i}^{\alpha}(s),X_{P_k}^\mu(s),s)ds + \int_{0}^{t}g_{P_k}(X_{A_i}^{\alpha}(s),X_{P_k}^\mu(s),s)dW_{P_k}(s).$$
Here, $(W_{P_k})_k$ is a sequence of independent standard Brownian motions. Similarly, $Y_{P_k}^{\mu}$
can be interpreted as the arrival times of passive particles, and $\Phi_{P_k}^{\mu}(t)$ is the cumulative lost service capacity over $[0,T]$.

\subsection{Coupling mechanisms between active and passive populations}

The drift coefficients $f_{A_i}$ and $f_{P_k}$ in \eqref{active} and \eqref{passive} can be considered as functions describing the interactions between active and passive particles. The specific formulations of $f_{A_i}$ and $f_{P_k}$ can be defined differentially in various models (see, e.g., \cite{Thieu2020-3, Philipowski2016}). Similarly, the detailed formulation of diffusion terms $g_{A_i}$ and $g_{P_k}$ in \eqref{active} and \eqref{passive} are defined depending on different models. 

%Next, we highlight key steps in our model construction.
%\begin{itemize}
%	\item[i.] The active population of the system is described as the following process:
%	
%	\begin{align}\label{active-eqn}
%	X_{A_i}^\alpha(t) = X_{A_i}^\alpha(0) + Y_{A_i}^{\alpha}(t) + \Phi_{A_i}^{\alpha}(t),
%	\end{align}
%	where $X_{A_i}^\alpha(t)$ denotes the position of the customer $i$ belonging to the group of active customers at time $t\geq 0$. On the other hand, $Y_{A_i}^{\alpha}$
%	can be interpreted as the arrival times of active customers, $\Phi_{A_i}^{\alpha}(t)$ is the cumulative lost service capacity over $[0,t]$.
%	\item[ii.] The passive population of the system is described as the following process:
%	
%	\begin{align}\label{passive-eqn}
%	X_{P_j}^\mu(t) = X_{P_j}^\mu(0) + Y_{P_j}^{\mu}(t) + \Phi_{P_j}^{\mu}(t),
%	\end{align}
%	where $X_{A_i}^\mu(t)$ denotes the position of the customer $j$ belonging to the group of passive customers at time $t \geq 0$. Similarly, $Y_{P_j}^{\mu}$
%	can be interpreted as the arrival times of passive customers, $\Phi_{P_j}^{\mu}(t)$ is the cumulative lost service capacity over $[0,t]$. 
%\end{itemize}
This behavioral system in the context of operations research applications can be viewed as an $M/M/\omega/N$ queueing system with the following assumptions:
\begin{itemize}
	\item [a)] there are $N$ particles that request services (for example, healthcare systems),
	\item [b)] the maximum number of requests in the system equals $\omega$,
	\item [c)] the arrival times of active and passive particles are exponential i.i.d random variables with intensities $\alpha$ and $\mu$, respectively.
\end{itemize}
Note that the notation $M/M/\omega/N$ means that the queueing model contains $\omega$ service points and $N$ places for waiting, while $M$ represents Poisson (or another probabilistic) arrival process (i.e., exponential inter-arrival times see, e.g.,  \cite{Kalashnikov1994}).

Motivated by \cite{Pilipenko2014} (cf. Section 3.3), we consider a limit theorem for a model of queueing theory that leads us to a system of
reflected SDEs.
Using the central limit theorem, we have

\begin{align}\label{limit-1}
\frac{Y_{A_i}^{\alpha}(t) - \alpha t}{\sqrt{\alpha}} \to w_{A_i}(t) \text{ when } \alpha \to \infty,
\end{align}
and
\begin{align}\label{limit-2}
\frac{Y_{P_k}^{\mu}(t) - \mu t}{\sqrt{\mu}} \to w_{P_k}(t) \text{ when } \mu \to \infty
\end{align}
in $\Lambda$.
Similarly, we also have

\begin{align}\label{Skorohod-1}
\left(\frac{X_{A_i}^\alpha(t)}{\sqrt{\alpha}},\frac{\Phi_{A_i}^\alpha(t)}{\sqrt{\alpha}} \right) \to (\xi_{A_i}(t), \phi_{A_i}(t)) \text{ when } \alpha \to \infty,
\end{align}
and
\begin{align}\label{Skorohod-2}
\left(\frac{X_{P_k}^\mu(t)}{\sqrt{\mu}},\frac{\Phi_{P_k}^\mu(t)}{\sqrt{\mu}} \right) \to (\xi_{P_k}(t), \phi_{P_k}(t)) \text{ when } \mu \to \infty
\end{align}
in $\Lambda$.

As a result, the pairs $(\xi_{A_i}(t), \phi_{A_i}(t))$ and $(\xi_{P_k}(t), \phi_{P_k}(t))$ are the solutions of the following Skorokhod type equations:

\begin{align}
\xi_{A_i}(t) &=  w_{A_i}(t) + \phi_{A_i}(t), \\
\xi_{P_k}(t) &=  w_{P_k}(t) + \phi_{P_k}(t).
\end{align} 

%In the lattice-based numerical implementation of this model, each customer is represented as a particle. The resulting coupled particle system is interacting via the site exclusion principle, i.e. each site of the lattice can be occupied only by a single particle.

%Note that an analogue of \eqref{limit-1}-\eqref{Skorohod-2} is valid not only for a Poisson process of arrivals or processing. Indeed, the functional central limit theorem (or Donsker is invariance principle) holds under various assumptions,
%see e.g. \cite{Billingsley1999}.

The reflecting stochastic queueing theory model we discussed here is a low-fidelity modelling methodology for general behavioral systems. In order to approximate this reflecting stochastic dynamics problem, we are going to combine the coupled system of SDEs with a high-fidelity modelling method by using a statistical-mechanics-based lattice gas approach. We provide more detailed descriptions of the lattice gas approximation by showing numerical examples in the next section.  

\section{Numerical examples}

In this section we consider an application of our general queueing theory model in a healthcare system. In particular, we investigate an interacting particles system of active and passive populations on the example of modelling the interactions of asymptomatic carriers and a susceptible population. We assume that active particles can be seen as asymptomatic hosts (carriers) while passive particles can be considered as susceptible individuals. The process is described as follows: the asymptomatic hosts do not know that they carry the virus and unknowingly spread the virus to the susceptible individuals. After that, the whole infected population may have severe symptoms and may need intensive care. Since the asymptomatic groups have not had any pronounced symptoms yet, the infectious dose of viable SARS-CoV-2 virus that needs to be present in the body to cause infection in another person is not known. Hence, we consider the case when a susceptible individual will be infected provided that individual communicates with asymptomatic carriers at his nearest neighbor sites.

%\section{Numerical examples}\label{KMC}
The representative examples we consider here for two different species of
\emph{active} and \emph{passive} particles, moving inside lattice $\Lambda$ defined as in the previous section
(we use in the notation the symbols A and P to 
respectively refer to them). Similarly, there are $\omega$ service gates at the service station and these service gates are located on the top row of the room. 
Note that the sites 
of the external boundary of the room, i.e. the sites  
$x\in\mathbb{Z}^2\setminus\Lambda$, are such that there exists 
$y\in\Lambda$ nearest neighbor of $x$ which
cannot be accessed by the particles. 
We call the state of the system
\emph{configuration} $\eta\in\Omega=\{-1,0,1\}^\Lambda$ 
and 
we shall say that the site $x$ is 
\emph{empty} if $\eta_x=0$,
\emph{occupied by an active particle} if $\eta_x=1$,
and
\emph{occupied by a passive particle} if $\eta_x=-1$.
The number of active (respectively, passive) 
particles in the configuration $\eta$ 
is given by 
$n_{\text{A}}(\eta)=\sum_{x\in\Lambda}\delta_{1,\eta_x}$
(resp.\ $n_{\text{P}}(\eta)=\sum_{x\in\Lambda}\delta_{-1,\eta_x}$), 
where $\delta_{\cdot,\cdot}$ is Kronecker's symbol.
Their sum is the total number of particles in the configuration $\eta$. 
\begin{figure}[h!]
	%\begin{figure}
	\centering
	\includegraphics[width=0.45\textwidth]{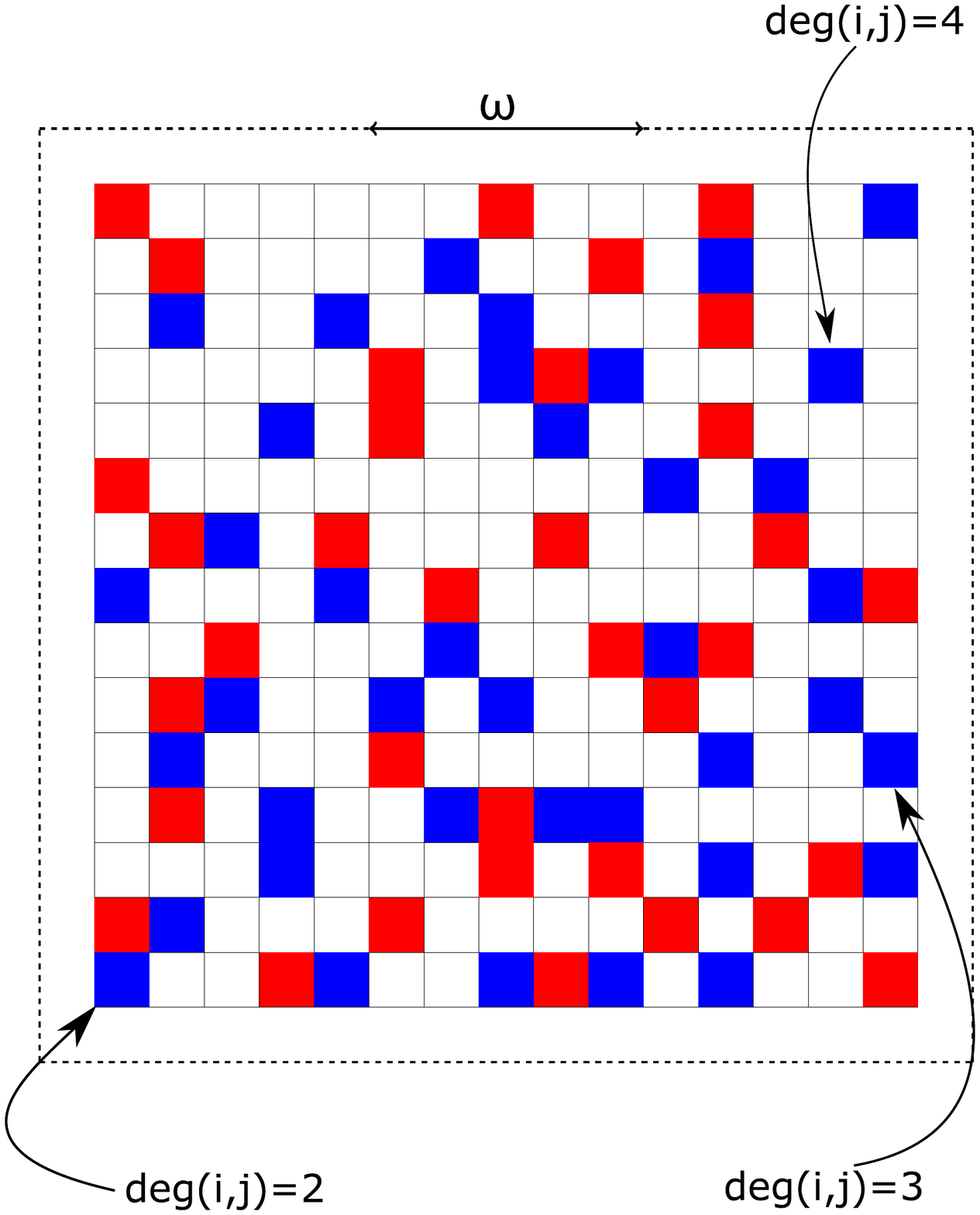}
	\vspace*{-18mm}
	\caption{(Color online) Schematic representation of our lattice model. Blue and red 
		squares denote passive and active particles, respectively, while the white squares within the geometry represents the empty spots. The thick dashed line surrounding a large
		fraction of the grid denotes the presence of reflecting boundary conditions. The service gates are located in presence of the arrow, with its width equal to $\omega$.}
	\label{fig:fig0}
\end{figure}

The original idea of this population dynamics has been proposed in \cite{Cirillo2019, Colangeli2019} in a lattice gas framework. Taking the inspiration from this stochastic dynamics in \cite{Cirillo2019}, we model the dynamics of our active and passive particles. Using the descriptions based on a simple exclusion process, the dynamics in the room is modeled as follows: the passive particles perform a 
symmetric simple exclusion dynamics on the whole lattice, whereas active particles are subject to a drift, guiding particles towards the service gates $\omega$. Here, active particles (infected individuals) pass through the service gates to enter the healthcare system since they need intensive care units. In general, most people who are infected with the COVID-19 virus will get mild to moderate respiratory illness and recover without requiring special treatment. Older individuals and also those groups of people of any age who have other serious health problems like cardiovascular disease, diabetes, chronic respiratory disease, and cancer are more likely to develop serious illness. We assume that the drift quantity $\varepsilon>0$ is the level of serious illness patients who need intensive cares. In particular, the larger $\varepsilon$ of active particles the more patients need intensive care units (the more load will be placed on the intensive care unit). Note that our dynamics are incorporated in the continuous time Markov chain $\eta(t)$ on $\Omega$ with rates $c(\eta,\eta')$, for the detailed definitions of the rates $c(\eta(t),\eta)$, see, e.g., \cite{Cirillo2019, Cirillo2020}. 

In this section, the numerical results are obtained by using the kinetic Monte Carlo (KMC) method. In particular,
we 
simulate the presented model by using the 
following scheme: we 
extract at time $t$ an exponential random time $\tau$ as a function with parameter which is 
the total rate 
$\sum_{\zeta\in\Omega}c(\eta(t),\zeta)$ of exchanging between configurations (e.g. in  \cite{Cirillo2019}), then set the time $t$ equal to $t+\tau$.
Next, we select a configuration using the probability 
distribution 
$c(\eta(t),\eta)/\sum_{\zeta\in\Omega}c(\eta(t),\zeta)$
and set $\eta(t+\tau)=\eta$ (for the detailed definitions of the rates $c(\eta(t),\eta)$, see e.g. \cite{Cirillo2019, Cirillo2020}).  

This numerical scheme has been studied in \cite{Cirillo2019}, where the authors implemented a version of KMC in the context of a lattice gas model with two species of particles. Further validation of the numerical methodology used here was reported in \cite{Cirillo2020}, based on a counterflow crowd dynamics model. The overall dynamics of our system is based on a continuous time Markov chain, i.e. the process will change state according to an exponential random variable and then move to a different state as specified by the probabilities of a stochastic matrix, together with a simple exclusion process. On the other hand, we use a statistical-mechanics-based lattice gas framework where we employ a kinetic Monte Carlo procedure to supplement the methodology described in the previous section. Furthermore, in general, Monte Carlo statistical methods, particularly those based on Markov chains, are one way to sample the distribution of the input random variable. Hence, the methodology used in our model can be considered as a multi-fidelity approach to statistical inference (e.g. in \cite{Peherstorfer2018-2,Robert2004}).

\begin{figure}[h!]
	\centering
	\begin{tabular}{lll}
		\includegraphics[width=0.33\textwidth]{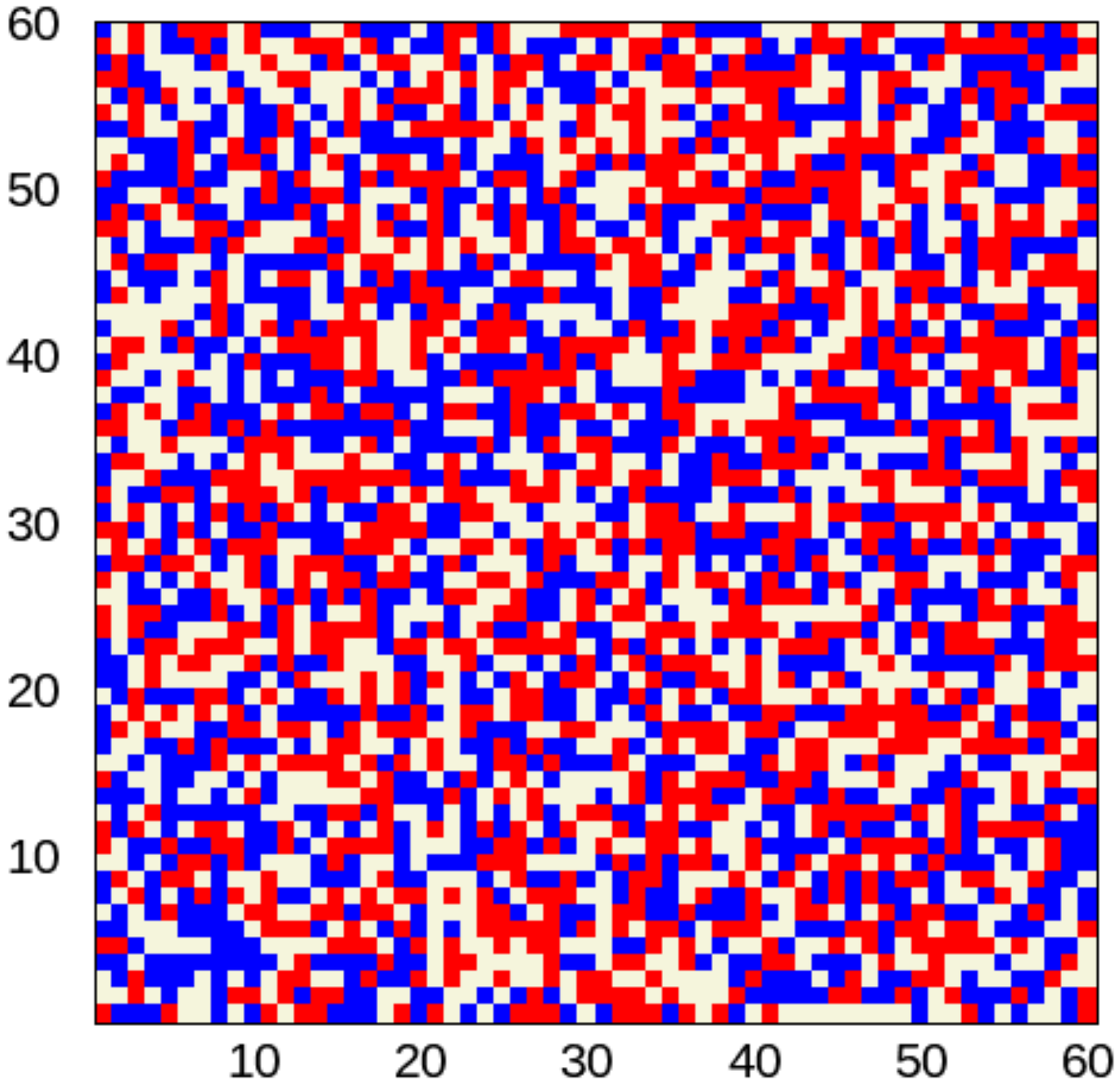}&
		\includegraphics[width=0.33\textwidth]{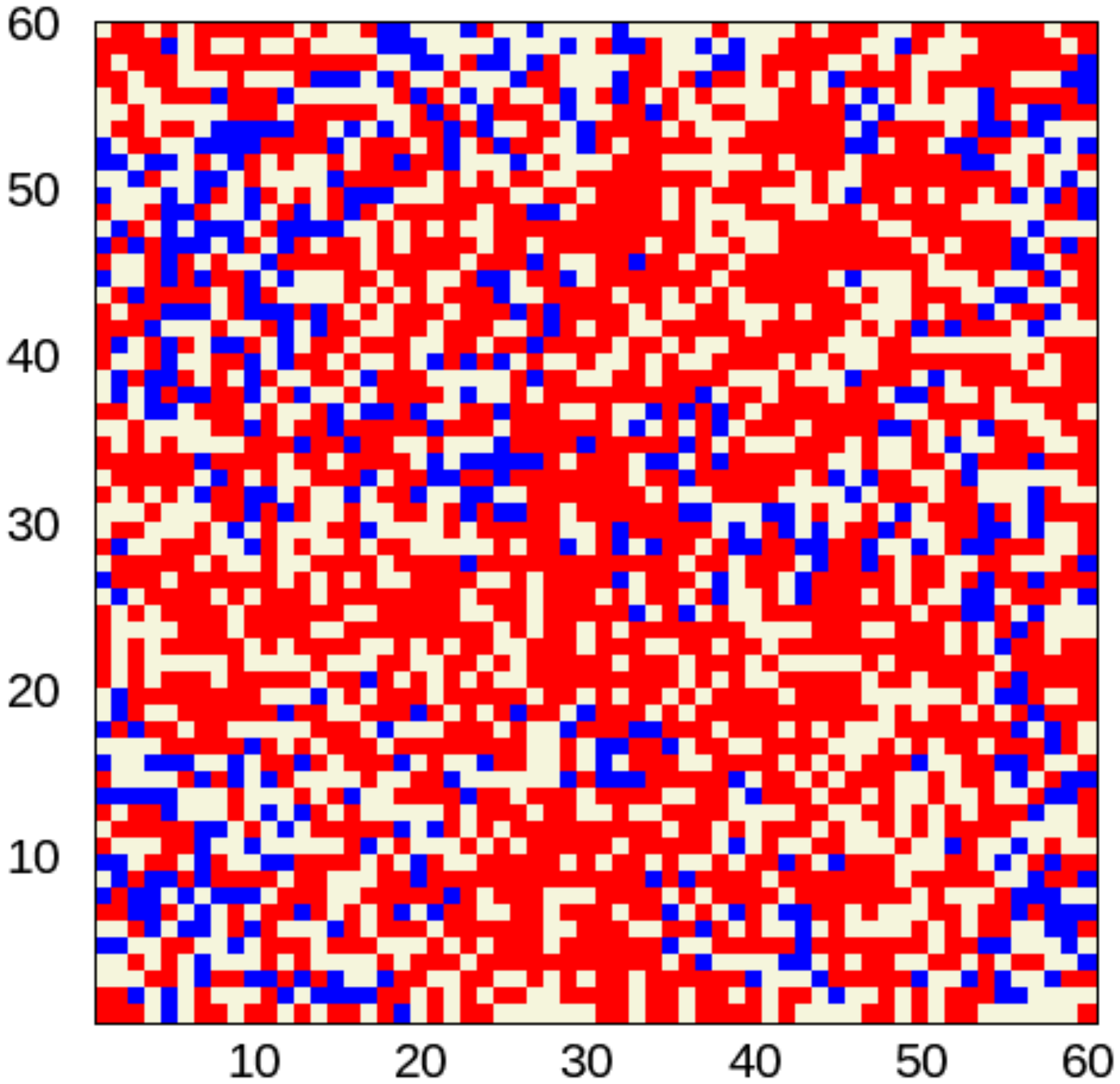}&
		\includegraphics[width=0.33\textwidth]{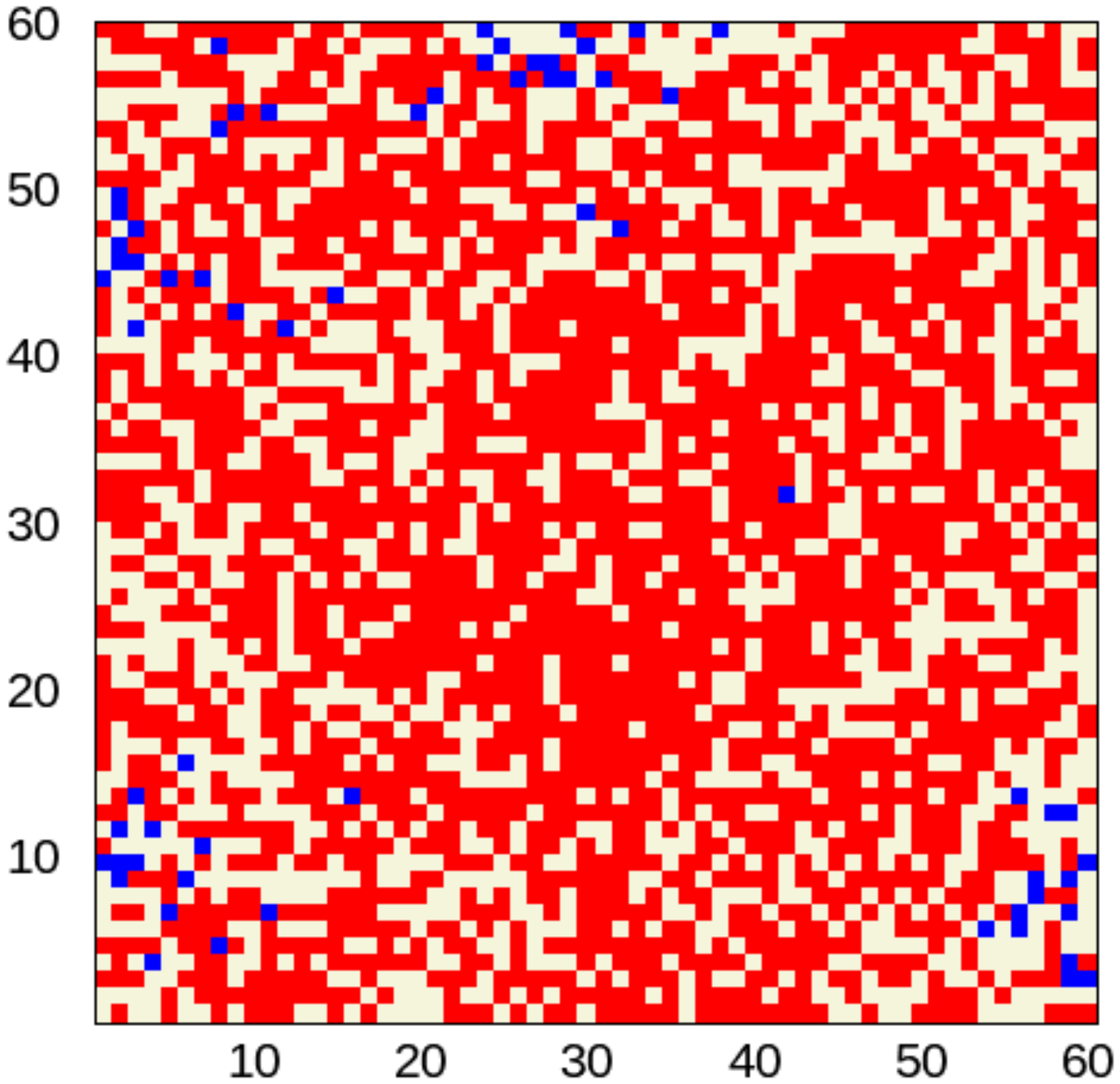}\\[0.1cm]
		\includegraphics[width=0.33\textwidth]{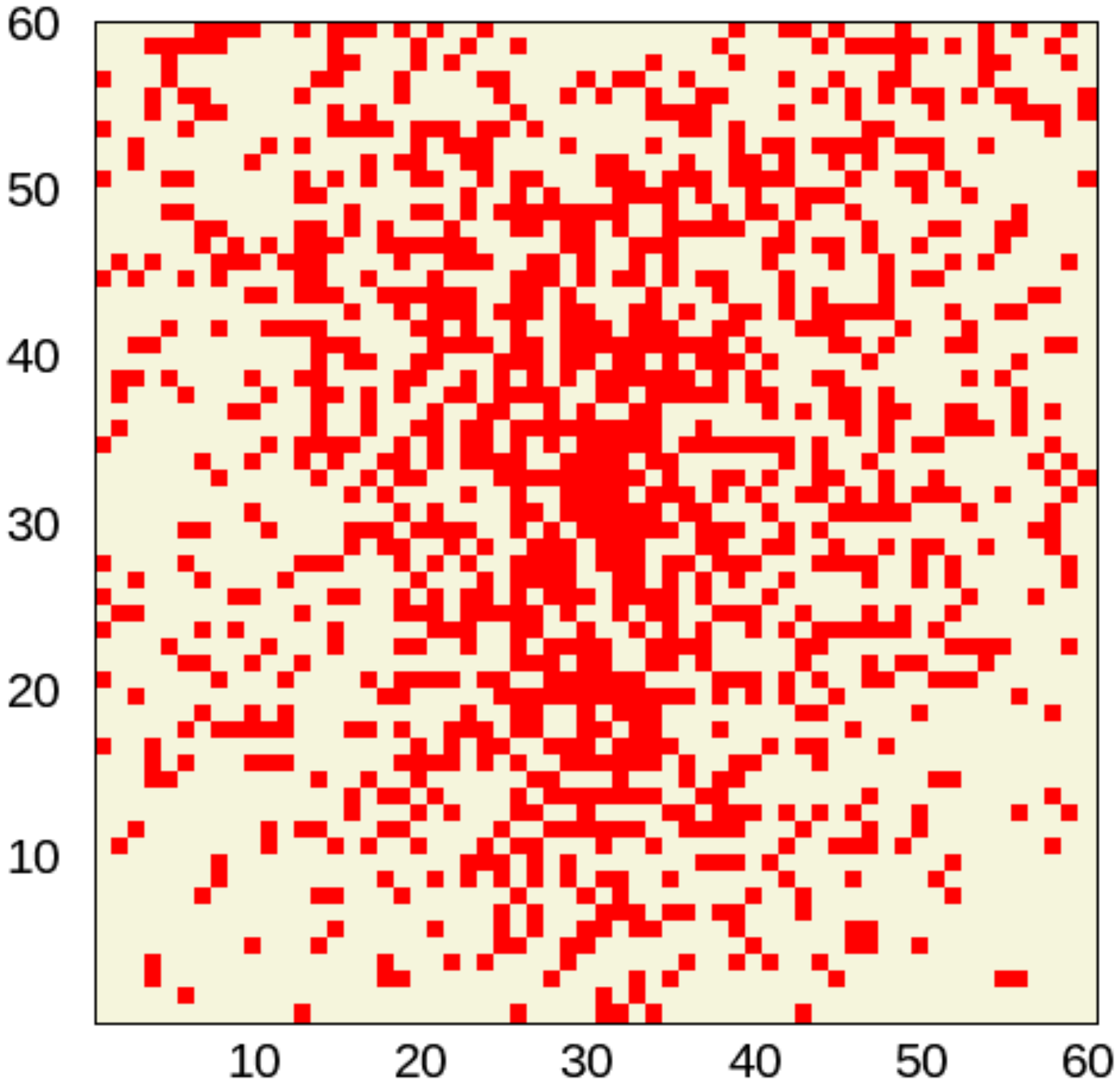}&
		\includegraphics[width=0.33\textwidth]{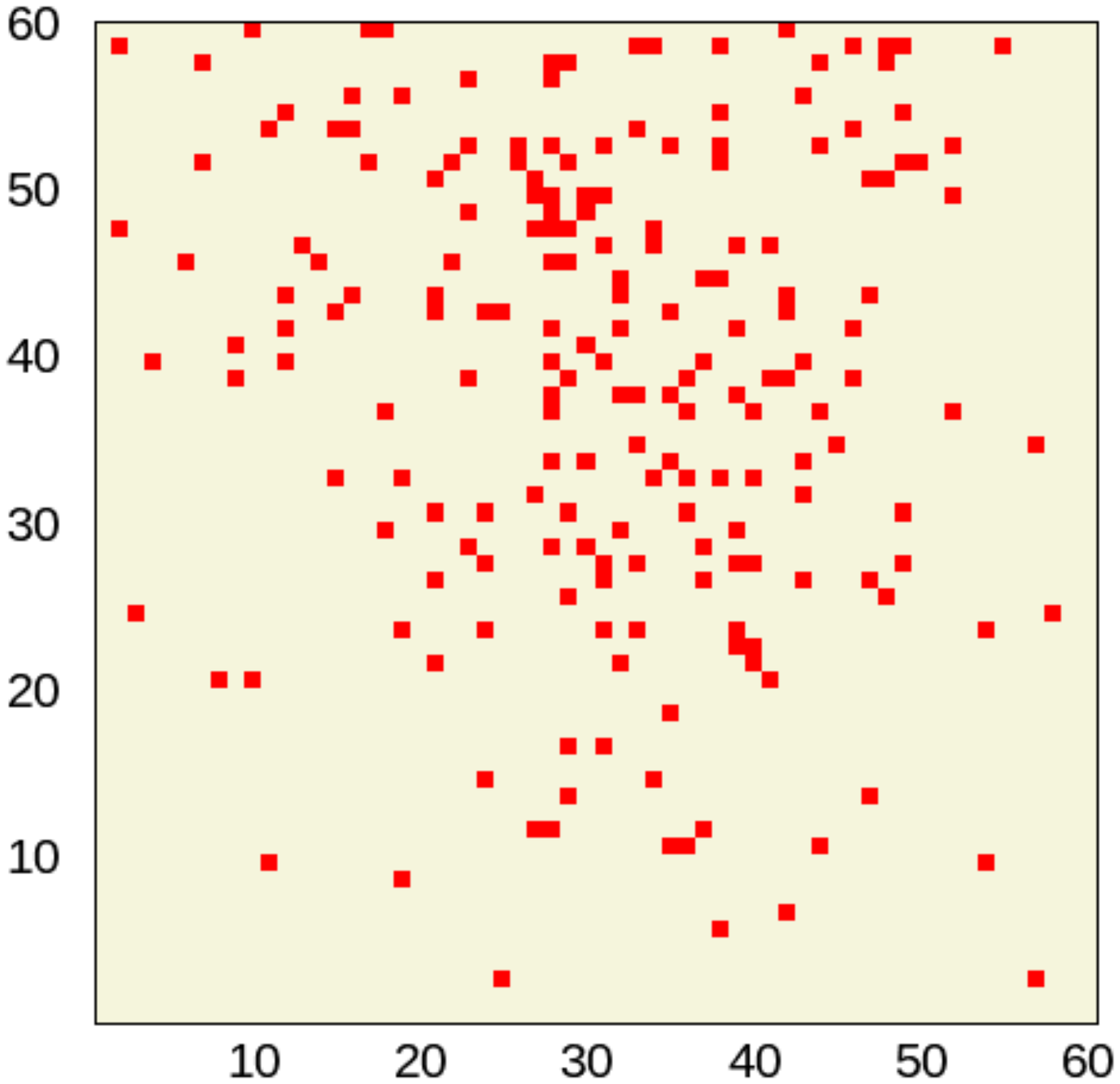}&
		\includegraphics[width=0.33\textwidth]{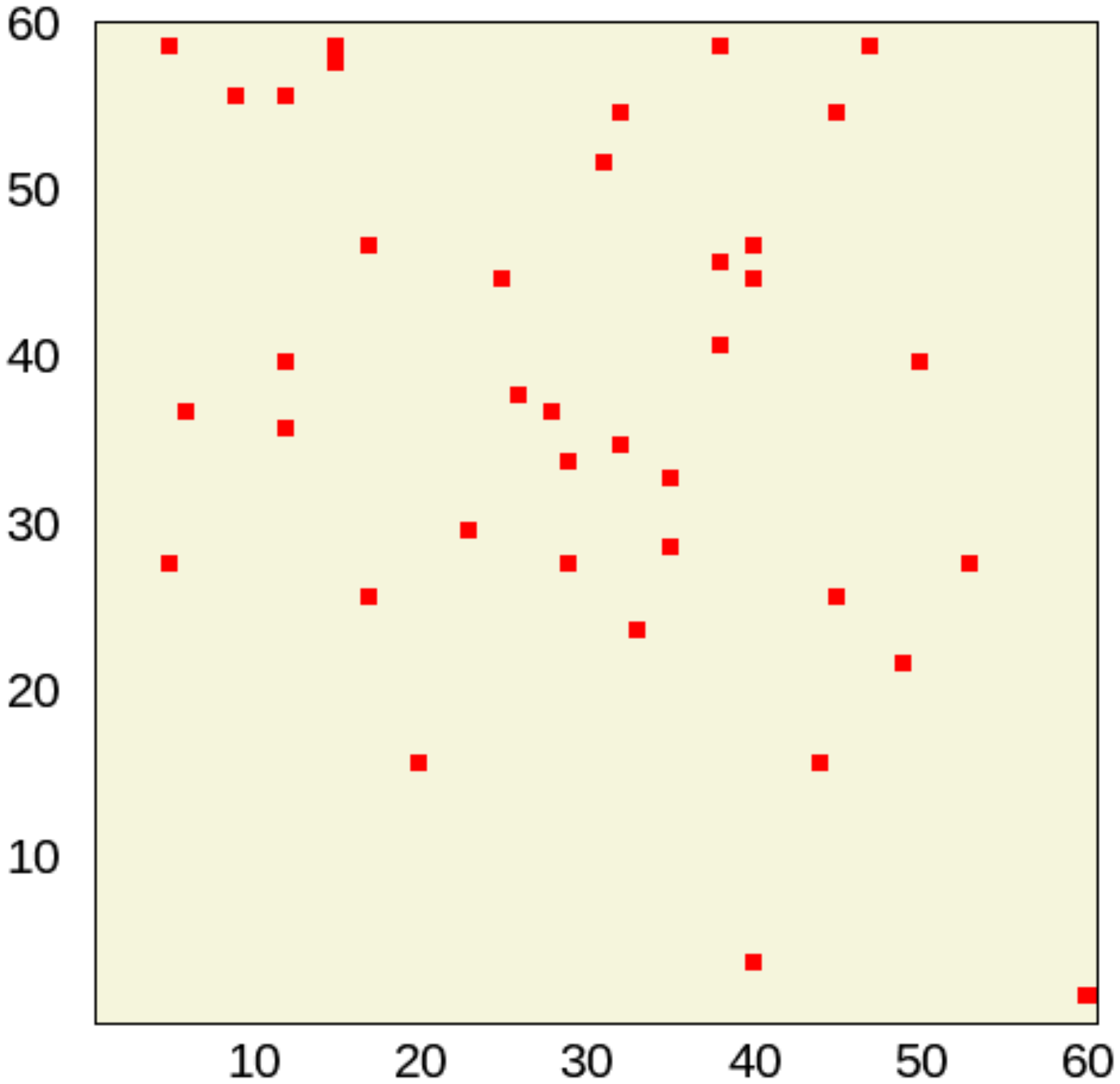}	
	\end{tabular}
	\caption{\small (Color online) The configurations of the model at different 
		times (increasing in lexicographic order).
		Parameters: $L=60$, $w_{\mathrm{ex}}=20$ and $\varepsilon=0.2$. 
		Red pixels represent active particles, blue pixels denote passive 
		particles, and gray sites are empty. 
		In the initial configuration (top left panel) there are $1200$ active and 
		$1200$ passive particles.}
	\label{fig:fig2}
\end{figure}

In the current version, we have improved the original simulations provided in \cite{Cirillo2019, Cirillo2020} by creating the interaction among active and passive particles. We have each site on the lattice connected by edges. The number of connectivities between these edges at a site $(i,j)$ will be called degree of that site, namely, $\textrm{deg}(i,j)$, see in Fig. \ref{fig:fig0}. The model is described as follows: if an arrival site $(i,j)$ on the lattice is occupied by a passive particle, then we count the number of active particles at its nearest neighbor sites. If the total number of actives particles at its nearest neighbor sites equals to $\textrm{deg}(i,j)-1$, then the passive particle switches to the active one (this is done by replacing the occupation number of the passive particle at its site equal to $0$, at the same time we let the occupation number of active particle to be equal to $1$ at the same site). The overall picture of our stochastic dynamics is illustrated in Fig. \ref{fig:fig2}, where we show the configuration of the system at different times. 

\begin{figure}[h!]
	\centering
	\begin{tabular}{ll}
		\includegraphics[width=0.43\textwidth]{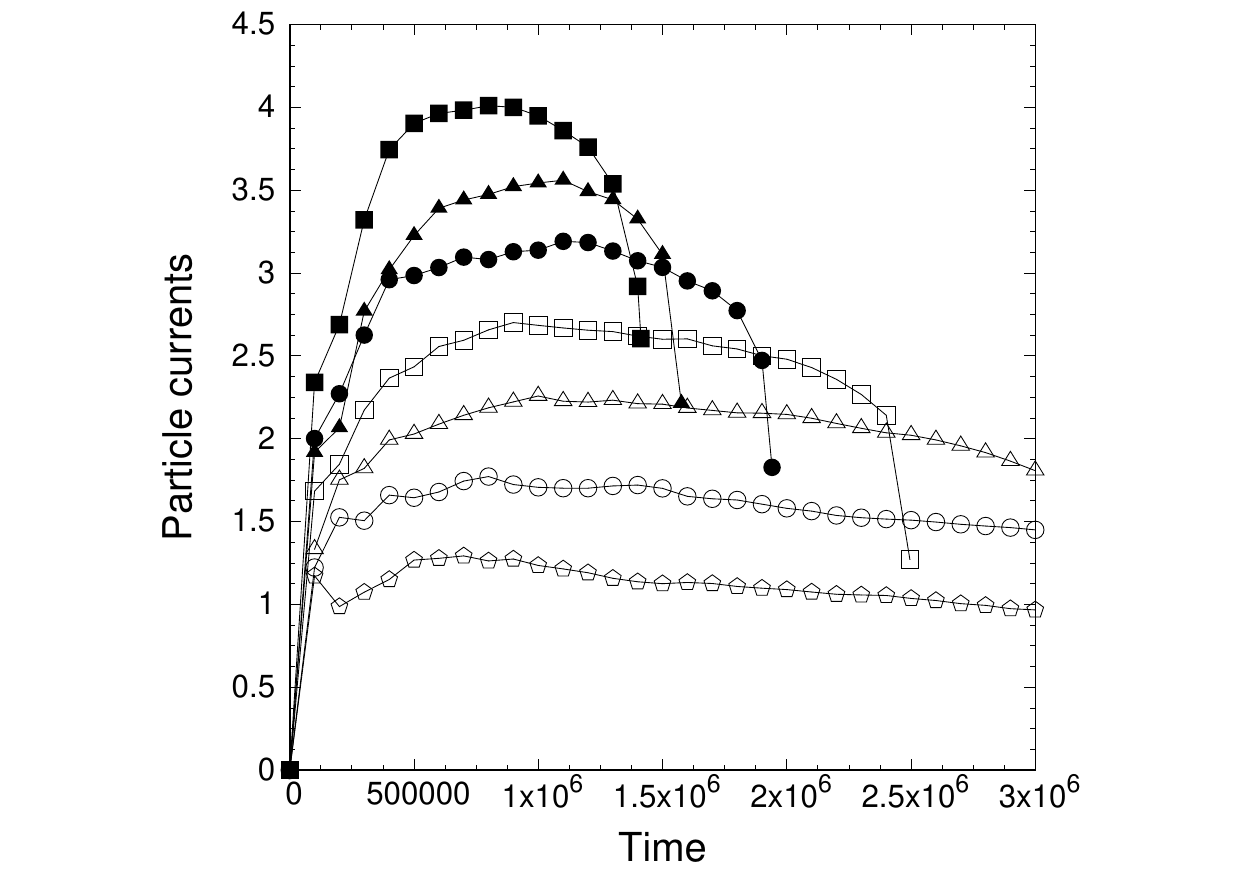}&
		\includegraphics[width=0.43\textwidth]{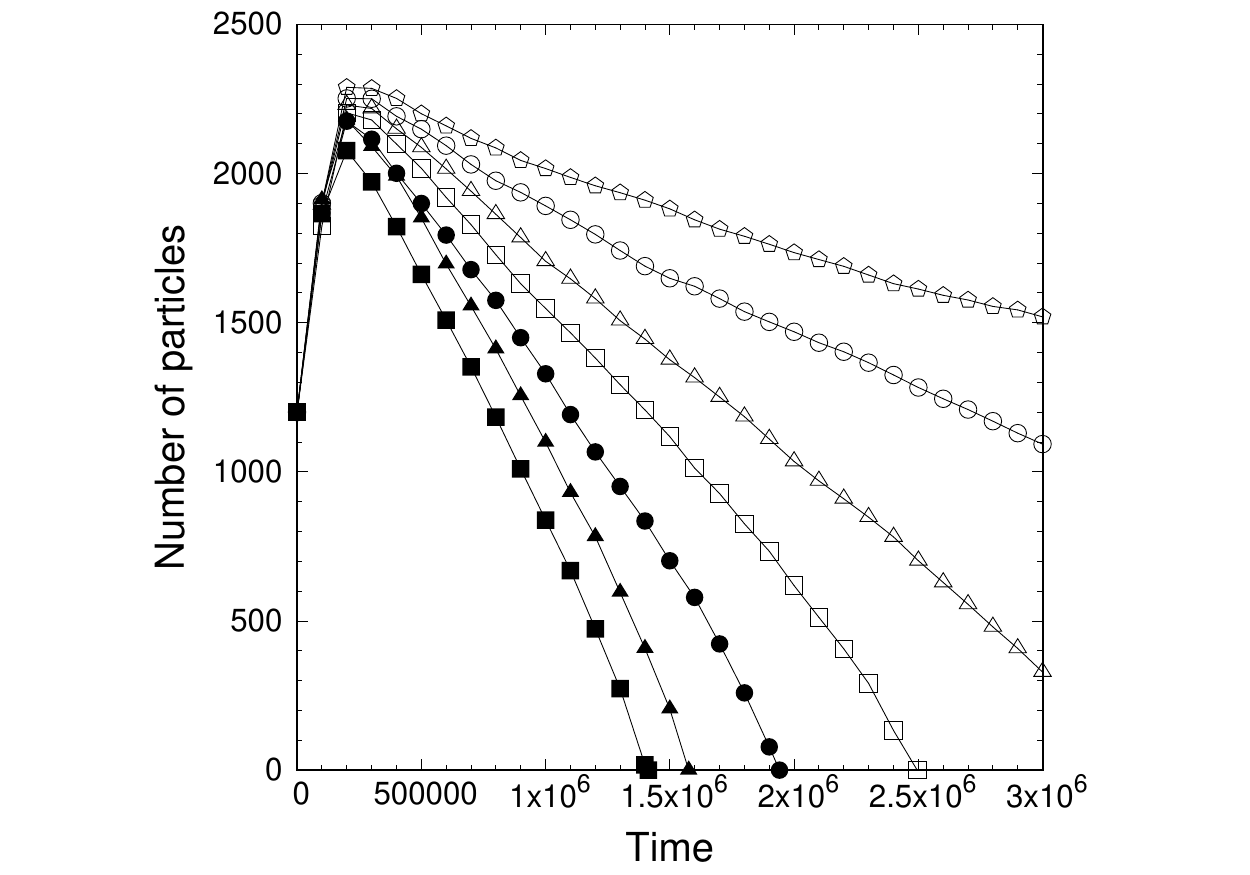}
	\end{tabular}
	\caption{Left panel: Evolution of the infected individual currents as a function of time for $L=60$, $N_{A}=N_{P}=1200$, $\varepsilon = 0$ (empty pentagons), $\varepsilon = 0.05$ (empty circles), $\varepsilon = 0.1$ (empty triangles), $\varepsilon = 0.15$ (empty squares), $\varepsilon = 0.2$ (solid circles), $\varepsilon = 0.25$ (solid triangles) and $\varepsilon = 0.3$ (solid squares). Right panel: Behavioral pattern of the infected individuals for $L=60$, $N_{A}=N_{P}=1200$,$\varepsilon = 0$ (empty pentagons), $\varepsilon = 0.05$ (empty circles), $\varepsilon = 0.1$ (empty triangles), $\varepsilon = 0.15$ (empty squares), $\varepsilon = 0.2$ (solid circles), $\varepsilon = 0.25$ (solid triangles) and $\varepsilon = 0.3$ (solid squares).}
	\label{fig:fig3}
\end{figure}

In the simulations, we fix the parameters for $L=30$, $\omega=20$ and $N_A=N_P=1200$.
The main numerical results of our analysis are shown in Fig. \ref{fig:fig3}, where we have plotted the total number of infected individuals and the infected individuals currents as functions of time, for different values of $\varepsilon$. The current is defined in the infinite time limit by the ratio between the total number of infected individuals, that in the interval $(0,t)$ passed through the service gates to enter the healthcare system, and the time $t$ (see e.g. in \cite{Colangeli2019}). Our numerical results have demonstrated that the latent infection transmissions caused by asymptomatic carriers can increase the needed intensive care. In Fig. \ref{fig:fig3}, the virus carriers increase over time. This is visible in all of the cases of $\varepsilon$. In the left panel of Fig. \ref{fig:fig3}, when we increase the values of $\varepsilon$ from $0$ to $0.3$, the currents of infected individuals increase. In particular, the current of virus carriers who need intensive care in the case of $\varepsilon = 0$ is smaller than in the case of $\varepsilon = 0.05$. Similarly, the current of virus hosts in the case of $\varepsilon = 0.05$ is smaller than in the case of $\varepsilon = 0.1$, and etc. When the virus carriers do not need intensive cares, this means that $\varepsilon = 0$. This is due to the fact that the infected individuals may or may not develop symptoms, or even have slight symptoms. The right panel of Fig. \ref{fig:fig3} shows the corresponding behavior of the total number of particles. It is clear that the number of virus carriers increases over time up to some value where it attains a maximum (at $N=2400$). However, soon after the number of virus hosts reaches its maximum, the number of infected individuals is reduced over time since they already entered the healthcare system by passing through the healthcare service gates. Furthermore, when $\varepsilon$ increases, there is an increase in the needed intensive care units of virus carriers. It is clear that if the asymptomatic populations spread the virus in the society, the virus carriers will significantly increase after short time. This effect is visible in the right panel of Fig. \ref{fig:fig3}, from $t=0$ to $t\approx 300000$ ($t$ represents the Monte Carlo time steps), the number of infected individuals dramatically increases from $1200$ to around $2300$. For example, assuming that the intensive care units capacity is $1200$, the hospital would not have enough beds for intensive care patients. Finally, it is worth noting that the effect of asymptomatic hosts can cause high pressure on the healthcare systems. To prevent this situation, the social lockdown and/or vaccination will need to be considered and built into the modelling scenarios. 

\section{Concluding remarks}
We have presented conceptually a procedure for combining coupled Skorokhod SDEs and a lattice gas framework for multi-fidelity modelling of complex behavioral systems. In particular, we have studied a coupled system of Skorokhod-type SDEs, modelling the interactions between active and passive populations. We have discussed the relationship between a general queueing theory model and a system of reflected SDEs via a limit theorem. 
%Using a statistical-mechanics-based lattice gas model, we approximated the interacting particles system by employing a kinetic Monte Carlo methodology. 
Using a multi-fidelity approach to statistical inference, we have considered an application of our reflected stochastic dynamics in the healthcare system, subjected to an epidemic, where the main focus in our analysis has been given to the interactions between asymptomatic and susceptible populations in a statistical-mechanics-based lattice gas framework. Based on our numerical experiments, we have observed that the virus spreading of asymptomatic hosts affects the healthcare system by an increase of intensive care units. The overall dynamics of our model is based on a continuous time Markov chain together with a simple exclusion process. In general, Monte Carlo methods, particularly those based on Markov chains, are now an essential component of the standard set of techniques used by quantitative researchers such as computational biologists, probabilists, and statisticians. In fact, the infection transmission of asymptomatic population is hard to measure.  Our multi-fidelity model has given a prediction on how the infection transmission of asymptomatic carriers behaves so that we can have suitable solutions to prevent the lack of intensive care units. 
In the current consideration, we have considered only the simplest case of the influence of asymptomatic carriers in the system. It would be instructive to analyze and  compare our current model with extensions of the model accounting for lockdowns and/or vaccination scenarios. Also, the comparison between the theoretical prediction of our model with the real data would need to be investigated further. Furthermore, the ideas presented in this contribution may be extended to other fields such as crowd dynamics and dynamics of neurons in nervous systems.
\section*{Acknowledgment}
{\small
Authors are grateful to the NSERC and the CRC Program for their
support. RM is also acknowledging support of the BERC 2018-2021 program and Spanish Ministry of Science, Innovation and Universities through the Agencia Estatal de Investigacion (AEI) BCAM Severo Ochoa excellence accreditation SEV-2017-0718 and the Basque Government fund AI in BCAM EXP. 2019/00432.
Both authors are acknowledging useful insight provided by Dr. Pilipenko, and
TKTT thanks Dr. A. Muntean (Karlstad, Sweden), Dr. E.N.M. Cirillo (Rome, Italy) and Dr. M. Colangeli (L'Aquila, Italy) for very fruitful discussions on the topic of active-passive populations dynamics.}
%\bibliographystyle{apasoft}
%\bibliography{mybibn}

\begin{thebibliography}{99}
%\bibitem{Zienkiewicz}  Zienkiewicz, O.C. and  Taylor, R.L. \textit{The finite element method}. McGraw Hill,
%Vol. I., (1989), Vol. II., (1991).
%\bibitem{Idelsohn} Idelsohn, S.R. and O\~{n}ate, E. Finite element and finite volumes. Two good friends.
%\textit{Int. J. Num. Meth. Engng.} (1994) \textbf{37}:3323--3341.
\bibitem{Bellomo2017} Bellomo, N., Bellouquid, A., Gibelli, L. and Outada, N. Mathematical models of crowd dynamics in complex venues.
\textit{Springer} (2017):119--160.
\bibitem{Skorohod1961} Skorohod, A.V. Stochastic equations for diffusion process in a bounded domain. \textit{Theory of Probability and Its Applications.} (1961) \textbf{VI}:264--274.

\bibitem{Zhang2016} Zhang, T. Lattice approximations of reflected stochastic partial differential equations driven by space–time white noise. \textit{The Annals of Applied Probability.} (2016) \textbf{26}:3602--3629.

\bibitem{Buisson2020} Buisson, J.D. and Touchette, H. Dynamical large deviations of reflected diffusions. \textit{Physical Review E}. (2019) \textbf{17}:460--477.

\bibitem{Thieu2020-3} Thieu, T.K.T. and Muntean, A. Well-posedness of a coupled system of {S}korohod-like stochastic differential equations. \textit{Submitted (arXiv:2006.00232)}. (2020).

\bibitem{Meares2020} Meares, H.D. and Jones, M.P. When a system breaks: queueing theory model of intensive care bed needs during the {COVID-19} pandemic. \textit{Med. J. Aust.} (2019) \textbf{212}:470--471.


\bibitem{Briand2020}  Briand, P., Grannoum, A., Labart, C. Mean reflected stochastic differential equations with jumps. \textit{Advances in Applied Probability}. (2020) \textbf{52}:523--562.

\bibitem{Schweitzer2019} Schweitzer, F. An agent-based framework of active matter with applications in biological and social systems. \textit{European Journal of Physics.} (2019) \textbf{40}:014003.

\bibitem{Zhang2020} Zhang, Y., You, C., Cai, Z., Sun, J., Hu, W. and Zhou, Z.H. Prediction of the {COVID-19} outbreak in {C}hina based on a new stochastic dynamic model. \textit{Scientific Reports.} (2020) \textbf{10}:21522.

\bibitem{Tadic2020}  Tad\'{i}c, B. and  Melnik, R. Modeling latent infection transmissions through biosocial stochastic dynamics. \textit{PLoS ONE}. (2020) \textbf{15}:e0241163.

\bibitem{Tadic2021}  Tad\'{i}c, B. and  Melnik, R. Microscopic dynamics modeling unravels the role of
asymptomatic virus carriers in SARS-CoV-2 epidemics at the interplay between biological and social factors. \textit{Computers in Biology and Medicine}. (2021) \textbf{133}:104422.

\bibitem{Subramanian2021} Subramanian, R., He, Q. and Pascual, M. Quantifying asymptomatic infection and transmission of {COVID-19} in {N}ew {Y}ork {C}ity using observed cases, serology, and testing capacity. \textit{PNAS.} (2021) \textbf{118}:e2019716118.

\bibitem{Peherstorfer2017} Peherstorfer, B., Kramer, B. and Willcox, K. Combining multiple surrogate models to accelerate failure probability estimation with expensive high-fidelity models. \textit{Journal of Computational Physics} (2017) \textbf{341}:61--75.

\bibitem{Peherstorfer2018-2} Peherstorfer, B., Willcox, K. and Gunzburger, M. Survey of multifidelity methods in uncertainty propagation, inference, and optimization. \textit{SIAM Review.} (2018) \textbf{60}:550--591.

\bibitem{Situ2005}  Situ, R. \textit{Theory of Stochastic Differential Equations with Jumps and Applications: Mathematical and Analytical Techniques with Applications to Engineering}. Springer (2005).
\bibitem{Philipowski2016} Philipowski, R. Interacting diffusions approximating the porous medium equation and propagation of chaos. \textit{Stochastic Processes and their Applications.} (2018) \textbf{117}:526--538.

\bibitem{Kalashnikov1994}  Kalashnikov, V.V. \textit{Mathematical Methods in Queuing Theory}. Springer Science Business Media Dordrecht (1989).

\bibitem{Pilipenko2014}  Pilipenko, A. \textit{An Introduction to Stochastic Differential Equations with Reflection}. Lectures in Pure and Applied Mathematics. Institutional Repository of the University of Potsdam, Germany (2014).
\bibitem{Cirillo2019} Cirillo, E.N.M., Colangeli, M., Muntean, A. and Thieu, T.K.T. A lattice model for active-passive pedestrian dynamics: a quest for drafting effects. \textit{Mathematical Biosciences and Engineering}. (2019) \textbf{17}:460--477.

\bibitem{Cirillo2020} Cirillo, E.N.M., Colangeli, M., Muntean, A. and Thieu, T.K.T. When diffusion faces drift: consequences of exclusion processes for bi--directional pedestrian flows. \textit{Physica D}. (2020) \textbf{413}:132651.

%\bibitem{Peherstorfer2018} Peherstorfer, B., Kramer, B. and Gunzburger, M. Preconditioning of the Cross-Entropy Method for Rare Event Simulation and Failure Probability Estimation. \textit{SIAM Review.} (2018) \textbf{60}:550--591.

\bibitem{Colangeli2019} Colangeli, M., Muntean, A., Richardson, O., and Thieu, T.K.T. Modelling interactions between active and passive agents moving through heterogeneous environments. \textit{in G. Libelli, N. Bellomo (Eds), Crowd Dynamics, Modeling and Simulation in Science, Engineering and Technology, Boston, Birkhauser, Springer}. (2019) \textbf{1}:211--254.

\bibitem{Robert2004}  Robert, C., and Casella, G. \textit{Monte Carlo Statistical Methods}. Springer (2004).


\end{thebibliography}

\end{document}